\documentclass[iop]{emulateapj}

\usepackage{times,natbib,graphicx}

\shorttitle{Reapproaching the Spin Estimate of GX 339-4}
\shortauthors{Ludlam et al.}

\begin{document}

\title{Reapproaching the Spin Estimate of GX 339-4}
\author{R. M. Ludlam\altaffilmark{1},
J. M. Miller\altaffilmark{2}, E. M. Cackett\altaffilmark{1}}
\altaffiltext{1}{Department of Physics \& Astronomy, Wayne State University, 666 W. Hancock St., Detroit, MI 48201, USA}
\altaffiltext{2}{Department of Astronomy, University of Michigan, 1085 South University Ave, Ann Arbor, MI 48109-1107, USA}

\begin{abstract}
We systematically reanalyze two previous observations of the black hole (BH) GX 339-4 in the very high and intermediate state taken with $\emph{XMM-Newton}$ and $\emph{Suzaku}$. We utilize up-to-date data reduction procedures and implement the recently developed, self-consistent model for X-ray reflection and relativistic ray tracing, {\sc relxill}. In the very high and intermediate state, the rate of accretion is high and thus the disk remains close to the innermost stable circular orbit (ISCO). We require a common spin parameter and inclination when fitting the two observations since these parameters should remain constant across all states. This allows for the most accurate determination of the spin parameter of this galactic black hole binary from fitting the Fe K$\alpha$ emission line and provides a chance to test previous estimates. We find GX 339-4 to be consistent with a near maximally spinning black hole with a spin parameter  $a_{*}$ $>0.97$ with an inclination of $36 \pm 4$ degrees. This spin value is consistent with previous high estimates for this object. Further, if the inner disk is aligned with the binary inclination, this modest inclination returns a high black hole mass, but they need not be aligned. Additionally, we explore how the spin is correlated with the power of the jet emitted but find no correlation between the two. 
\end{abstract}

\section{Introduction}
GX 339-4 is an important source for accretion physics. It is a recurrent low mass X-ray binary system (LMXB) that has a prominent disk reflection spectrum. This galactic black hole binary (GBHB) has undergone four outbursts over the last twelve years, thus making it the one of the most active transient systems \citep{plant14}. Throughout its outbursts, GX 339-4 has been observed in the very high/soft state (VHS), intermediate state, and the low/hard state (LHS) followed by a very faint state, but has not yet been observed in the quiescent state \citep{hynes03}.  See \citet{McRem2006} for a review on states. The mass of the BH and inclination are determined through interactions with the companion star. Through the analysis of emission lines during an outburst, \citet{hynes03} determined a mass function of $5.8\pm0.5$ M$_\odot$, where the mass function can be expressed in the form of 
\small
\begin{equation}
f(M)=\frac{M_{x} \sin i^{3}}{(1+q)^2}
\end{equation}
\normalsize 
$M_{x}$ being the mass of the compact object, $q$ the mass ratio of the companion to the compact object, and $i$ the inclination.  The inclination of the BH is still under some speculation since it may deviate from the inclination of the system itself, which is discussed in further detail in \S 3. The distance was determined using the optical/IR absorption along the line of sight to GX 339-4. It was found to lie between 6-15 kpc \citep{hynes04} though this value is generally quoted as a distance of 8 kpc \citep{zd04}.

GX 339-4 is an ideal source for applying new models that use the method of fitting Fe K$\alpha$ emission to estimate spin because of its prominent reflection features. The advantage of using the iron emission line is that it eliminates the need to know the mass or distance to the object. The determination of mass and distance is complicated by large uncertainties for most objects. 

Models that allowed for the spin of a BH to be a variable parameter that could be constrained directly through the fitting of spectra emitted from an object were not available until 2004.  However, line models did give important early indications of very high spin parameters \citep{miller02}. Since then, there have been numerous additive and multiplicative models developed to describe the different features that appear in the high energy X-ray band. These include models such as {\sc relline} (\citealt{dauser10}, \citeyear{dauser13}) and {\sc xillver} (\citealt{garkall10}, \citeyear{garcia13}) among others. 
The latest of these models to be developed is the combination of the relativistic ray tracing kernel, {\sc relline}, and X-ray reflection code, {\sc xillver}, into one self consistent model {\sc relxill} \citep{garcia14}.

GX 339-4 was estimated to harbor a near maximally spinning BH during an outburst in 2002-2003 \citep{miller04b}. Observations were taken by $\emph{Chandra}$, $\emph{XMM-Newton}$, and the $\emph{Rossi X-Ray Timing Explorer (RXTE)}$. This observation was among the first to use burst mode aboard $\emph{XMM-Newton}$, and some aspects of the reduction and systematics were unknown.  \citet{Kirsch} later extensively went through analyzing burst mode data taken of the Crab Nebula in order to allow for a more conclusive extraction of spectra. 
In any case, \citet{miller04b} ruled out the possibility of this object being a non-spinning BH at more than the 8 $\sigma$ confidence level from the iron (Fe K$_\alpha$) emission line profile. They were able to estimate that the accretion disk extended in to $(2-3) R_{g}$, which translates to $a_{*} \geq 0.8-0.9$. \citet{reis08} reanalyzed the observations of GX 339-4 in the VHS and LHS (\citealt{miller04b}, \citeyear{miller06}) which were observed by $\emph{XMM-Newton}$. They used updated reflection models ({\sc{reflionx}} and {\sc{refhiden}}) that were able to take into account blackbody emission and Compton scattering in the accretion disk that were not available prior. The results of their extensive modeling was a spin of $0.935\pm0.001$ which supported the previous findings. 

These results were strengthened further by observations taken by $\emph{Suzaku}$ after an outburst that occurred during 2006-2007.  \citet{miller08} fit the spectra obtained during the outburst with two other spectra taken by XMM while GX 339-4 was in a higher and lower state. The joint fit across states returned a BH with a spin parameter of $a_{*}=0.93\pm0.01$. This was again among the first observations in a burst mode, and the joint fitting procedures served as an important check.
Later analysis of the $\emph{Suzaku}$ data by \citet{yamada} found that the spectra may be piled up; however, a comprehensive study of pile-up by \citet{miller10} shows that pile-up only acts to narrow reflection features, and cause spin to be under-estimated.

Spin may be hard to determine in the low/hard state as can be seen in \citet{dauser14} and \citet{Fabian14a}, therefore making it necessary to focus on the very high and intermediate states using new models to deliver the best possible spin measurement for GX 339-4. This focus on the VHS and intermediate state has to do with the location of the inner disk in relation to the ISCO. This is discussed further in \S 3.
We reanalyze the observations from \citealt{miller04b} and \citealt{miller08} using the latest data reduction procedures and model. This paper is formatted as follows: \S 2 describes the observations and data reduction process. \S 3 is the spectral analysis and results. \S 4 presents a discussion of the findings of this work.

\section{Observations and Data Reduction}
\subsection{$\emph{XMM-Newton}$}
We reanalyze the 75.6 ks observation of GX 339-4 in the very high state during revolution 0514 that began on 2002 September 29 at 09:06:42 UT. The EPIC-pn camera (\citealt{struder}) was operated in burst mode with a thin optical filter. The data were reduced using $\emph{XMM-Newton}$ Scientific Analysis System (SAS ver. 13.5.0). We also follow the guidelines for data reduction in burst mode from \citet{Kirsch}. The data were reduced two different ways using the task {\sc epfast} and again without the task since the use of {\sc epfast} may alter the energy scale of the Fe K band \citep{walton12}. RAW space was set to RAWY$\leq{140}$ and RAWX was 10 columns across (30-40). Above RAWY of 140 the determination of normalization becomes increasingly more affected based upon examination of figure two in Kirsch et. al. (2006). Although photon indices don't become affected until above RAWY$>160$, choosing to lower the limit ensures that $\Gamma$ is unaffected by pile up and a better determination of the normalization. In addition, the events were filtered to allow single and double events (PATTERN$\leq{4}$) while excluding bad pixels (FLAG==0). 

Since the object is so bright, there was not a region available to extract a background. 
 Most observations of GX 339-4 with $\emph{XMM-Newton}$ have operated the EPIC-pn camera in \lq \lq timing" mode, even when the source is not bright, serving to complicate background estimates.  An entire Epic-pn CCD is exposed in timing mode.  To estimate an upper limit on the background, we extracted an area equivalent to a Epic-pn CCD (4.4 by 13.6 arcmin) from the MOS-1 exposure of GX 339-4 obtained in obsID 0112900201.  We chose a region next to GX 339-4, and with a space density of background sources that is higher than average, as a kind of worst-case scenario.  Extracting the data from this region, including point sources and any diffuse emission, gives a flux of $f = 1.1\times 10^{-12}~ {\rm ergs}~ {\rm cm}^{-2}~ {\rm s}^{-1}$.  In fact, this limit is much too high, as the extraction region that we used within timing mode is smaller than the full CCD.  Even so, this flux is four orders of magnitude below the flux measured ($f = 1.1576 \times 10^{-8}~ {\rm ergs}~ {\rm cm}^{-2}~ {\rm s}^{-1}$) from GX 339-4 in the observation of interest. 

Response files were generated using SAS {\sc rmfgen} and {\sc arfgen} commands. There are at least 20 counts per energy bin to allow the use of $\chi^{2}$ minimization when performing spectral fits. GX 339-4 is determined to be in the very high state at the time of this observation due to the spectral parameters found in \citet{miller04b} and the timing properties found in \citet{homan05a}.
\subsection{$\emph{RXTE}$}
$\emph{RXTE}$ simultaneously observed GX 339-4 on 2002 September 29 at 09:12:11 UT for 9.6 ks. There was not a simultaneous observation taken at the time of the $\emph{Suzaku}$ observation, but rather, there were four other observations taken on the same day for 3.5, 3.41, 3.4, and 1.8 ks. We chose to use the standard products that are generated from standard modes after each observation as a guide for the photon index of the $\emph{XMM-Newton}$ and $\emph{Suzaku}$ observations.  We added $0.6\%$ systematic error to the spectra using the HEASOFT (ver. 6.15) sub package FTOOL {\sc grppha} task. The Proportional Counter Array (PCA; \citealt{jahoda}) spectrum is generated from the standard 2 mode which covers 129 energy channels and has a 16s time resolution. The two High Energy X-ray Timing Experiment (HEXTE; \citealt{rothschild}) spectra (from Cluster 0 and Cluster 1) have the same time resolution as the PCA and cover 129 spectral channels. Refer to $\emph{The RXTE Cook Book}$ for all these details. When performing spectral fits, the PCA was fit in the 3.0-25.0 keV energy range and HEXTE fit the 20.0-100.0 keV energy range.
\subsection{$\emph{Suzaku}$}
We reanalyze the observation of GX 339-4 in the intermediate state performed by $\emph{Suzaku}$ on 2007 February 12 beginning at 05:33:31 UT. This observation is classified as being in the intermediate state due to many properties that are consistent with previous results obtained in the intermediate states \citep{miller04a}. We processed the unfiltered event files through {\sc aepipeline} to produce clean event files for both the X-ray Imaging Spectrometer (XIS; \citealt{koy}) and Hard X-ray Telescope (HXD; \citealt{kokubun}and \citealt{tak07}) instruments. XIS was operated in 0.3s burst mode with the $1/4$ window option selected. HXD cameras were operated in the standard mode. All proper calibration products were determined and applied from the HEASARC Calibration Database (HXD20110913, XIS20120209, and XRT20110630). 

We chose to use the two working \lq front-illuminated' cameras, XIS0 and XIS3, in our analysis. The \lq back-illuminated'  instrument, XIS1, was not used due to calibration uncertainties as per \citet{Plant15}. New attitude files were created for both editing modes on each camera using {\sc aeattcor.sl}\footnote{http://space.mit.edu/ASC/software/suzaku/aeatt.html} to account for the wobbling that Suzaku undergoes. The attitude files were then applied to the clean event files by using the FTOOL {\sc xiscoord}. Even though the 1/4 window and burst option were selected for this observation, there is still a significant amount of photon pileup in the data. Pileup causes the photon index of a spectrum to appear harder due to the inability of the detector to register individual events. Two or more events will be interpreted as a single event with a higher energy (see \citealt{miller10} for a full description of the problem and consequences). Therefore, we use the RXTE observations that were taken on the same day as a guide for the photon index of GX 339-4 in the 2.0-10.0 keV energy range. We find that excluding data affected by pileup at the 8\% level produces the proper spectral slope. We employ the tool  {\sc pileest} to determine the contours for this region for XIS0 and XIS3. We then exclude this region when using XIS {\sc xselect} by using an annulus when extracting spectra. We used an outer radii of 7.5$^\prime$ with an inner radius of 1.1$^\prime$ and 1.4$^\prime$ for XIS0 and XIS3 respectively. There were not available regions in the images to extract a background. Response files were created for each spectrum using the task {\sc xisresp} before combining the two XIS spectra together with the FTOOL {\sc addascaspec}. Last, the combined spectrum was run through {\sc grppha} to require a minimum of 20 counts per energy bin like the $\emph{XMM-Newton}$ spectra.

The HXD/PIN was operated using the XIS nominal pointing. The proper response and non-X-ray background (NXB) files were obtained before executing the FTOOL {\sc hxdpinxbpi} on the cleaned event. This tool creates a source spectrum that is corrected for dead time as well as a combined background (cosmic and non-X-ray) spectrum. A minimum of 20 counts per energy bin was required again. The GSO cleaned event file was run through the FTOOL {\sc hxdgsoxbpi} after the appropriate NXB, spectral binning and additional ARF files were downloaded. The GSO NXB is already binned so the source spectrum must be binned accordingly. When performing spectral fits the XIS, PIN, and GSO were fit in the 1.0-10.0 keV, 15.0-60.0 keV, and 70.0-600.0 keV energy ranges respectively.

\section{Spectral Analysis and Results}
We use XSPEC version 12.8.1 \citep{arnaud96} in this work and errors are quoted at at least the 90 $\%$ confidence level. We utilize the new {\sc relxill} v0.1e modeling package \citep{garcia14} to properly describe the reflection and relativistic effects within the spectra, specifically to fit the Fe K$\alpha$ emission line and determine the spin of this object. We combine {\sc relxill} with the additive model {\sc diskbb} to account for the thermal component of the disk and the multiplicative model {\sc tbabs} to account for absorption along the line of sight. 
The absorption column is fixed throughout this analysis to be  $5.26 \times 10^{21}$ cm$^{-2}$ in accordance with the findings by \citet{dl90}. Note that GX 339-4 was in the \lq \lq very high" state during the $\emph{XMM-Newton}$ observation and in the intermediate state for the $\emph{Suzaku}$ observation. 

The inclination of GX 339-4 has not been determined conclusively. It is known that the orbital inclination has to be less than 60$^\circ$ from the lack of eclipses present in the optical data \citep{cowley02}. Futher, it can not have a inclination of much less than $\sim$ 40$^\circ$ in order to have a dynamical mass consistent with the findings of \citet{hynes03}. Any lower of an inclination would give a BH with a mass greater than 20 M$_\odot$. \citet{yamada} found that the whole 25$^\circ$-45$^\circ$ range is allowed at the 90\% confidence level. On the other hand, the orbital inclination may not even be aligned with the inclination of the BH itself (see \citealt{maccarone}). We choose to allow the inclination to be free when fitting the spectra though we find it to be consistent with the 90\% confidence range determined by \citet{yamada}.

\begin{table}
\caption{Joint Fitting of $\emph{Suzaku}$ and $\emph{XMM-Newton}$ with HXD/PIN}
\label{tab:joint} 
\begin{center}
\begin{tabular}{llcc}
\hline
Component & Parameter & $\emph{Suzaku}$ & $\emph{XMM-Newton}$\\
\hline
{\sc tbabs}
&$N_\mathit{H} (10^{22}) '$
&\multicolumn{2}{c}{0.526}
\\
{\sc diskbb}
&$T_\mathit{in} (keV)$
&$0.76 \pm 0.01$ 
&$0.80 \pm 0.01$ 
\\
&norm
&$2230_{-60}^{+110}$ 
&$1640_{-80}^{+90}$ 
\\
{\sc relxill}
&$q_{in}$
&$9.9_{-6.9}^{+0.1}$ 
&$7 \pm 1$ 
\\
&$q_{out}$
&$0.0002_{-0.0001}^{+0.9221}$ 
&$2.4_{-0.5}^{+0.7}$ 
\\
&$R_{break} (R_g)$
&$5.3_{-2.6}^{+0.4}$
&$4.4 \pm 1.5$
\\
&$a_{*}$
&\multicolumn{2}{c}{$>0.97$}
\\
&$\mathit{i} (^{\circ})$
&\multicolumn{2}{c}{$36 \pm 4$} 
\\
&$R_\mathit{in} (ISCO) '$
&\multicolumn{2}{c}{1}
\\
&$R_\mathit{out} (R_\mathit{g}) '$ 
&\multicolumn{2}{c}{400}
\\
&$\mathit{z} '$
&\multicolumn{2}{c}{0}
\\
&$\Gamma$
&$1.99\pm0.01$ 
&$2.56 \pm 0.02$ 
\\
&$log(\xi)$
&$4.36\pm0.04$ 
&$4.7_{-0.3}^{+1.3}$ 
\\
&$A_\mathit{Fe}$
&\multicolumn{2}{c}{$4.8_{-0.7}^{+2.0}$}
\\
&$E_\mathit{cut} (keV)$
&\multicolumn{2}{c}{$100_{-5}^{+9}$}
\\
&$\mathit{f}_\mathit{refl}$
&$2.4_{-1.7}^{+0.8}$ 
&$1.1_{-0.5}^{+1.7}$ 
\\
&angleon$'$
&\multicolumn{2}{c}{1}
\\
&norm
&$0.8 \pm 0.2$
&$0.9_{-0.4}^{+1.0}$ 
\\
\hline
&$\chi_\nu^{2}$(dof)
&\multicolumn{2}{c}{1.23 (1472)}
\\
\hline
$'$ = fixed
\end{tabular}

\medskip
Note.--- Errors are quoted at $\geq$ 90 \% confidence level. $N_{H}$ was fixed as the  \citet{dl90} value. A constant was allowed to float between the XIS and HXD/PIN spectra. The XIS was frozen at the value of 1.0 and the HXD/PIN factor was fit at 0.857. A Gaussian was modeled at 1.57 keV with an equivalent width of $6.38\times10^{-5}$ keV and $3.08\times10^{-2}$ normalization. The inner emissivity has been restricted between 3-10. The outer emissivity has been restricted between 0-3. The break radius was restricted between 3-6 $R_{g}$. $q_{in}$ for the $\emph{Suzaku}$ spectra is not well constrained within restricted boundaries. Setting angleon=1 takes the inclination into account when modeling reflection.  $log(\xi)$ for $\emph{XMM-Newton}$ is consistent with the hard limit of 6 within the 90\% confidence level.

\end{center}
\end{table}

\begin{figure}
\centering
\includegraphics[angle=270,width=8.4cm]{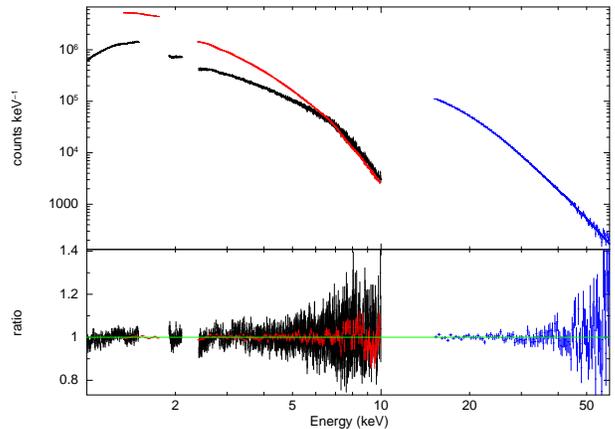}
\caption{Simultaneous fit of $\emph{Suzaku}$ XIS (black) and $\emph{XMM-Newton}$ (red) with the addition of the HXD/PIN spectra (blue) extending to the higher energies. XIS data has been rebinned for plotting purposes. See Table 4 for parameter values. See Figure 2 for unfolded model spectrum.}
\label{fig:joint}
\end{figure} 

\begin{figure}
\centering
\includegraphics[angle=270,width=8.4cm]{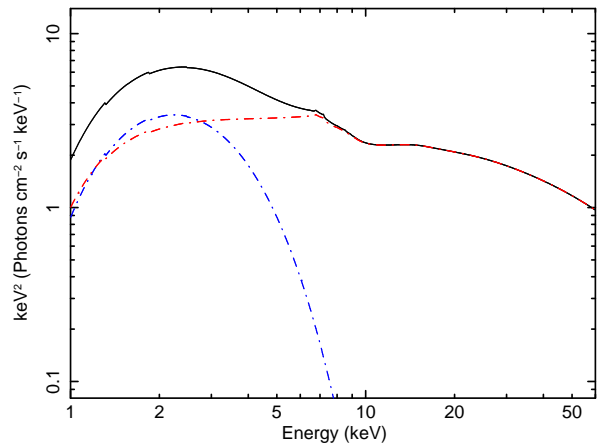}
\caption{Unfolded model spectrum from Figure 1 for $\emph{Suzaku}$  XIS and HXD/PIN. The blue component corresponds to {\sc diskbb}. The red component illustrates {\sc relxill}.}
\label{fig:unfolded}
\end{figure} 

\begin{figure}
\centering
\includegraphics[width=8.4cm]{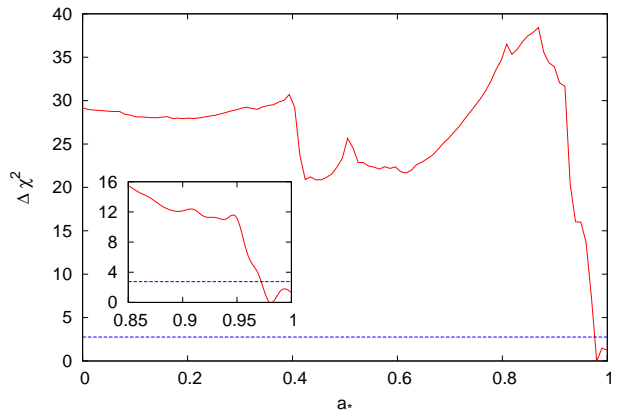}
\caption{The change in goodness-of-fit versus spin taken over 100 evenly spaced steps generated with XSPEC \lq \lq steppar". The spin was held constant at these steps while the other parameters were free to adjust. The inset shows a closer view of how $\chi^2$ changes for spin greater than 0.85. The blue dotted line in both cases is the 90\% confidence level. }
\label{fig:contour}
\end{figure} 

We fit the two observations jointly in order to obtain better constraints on the parameters of greatest interest. The individual spectra provide fits of comparable quality, but do not give equally strong constraints on their own. Tying the $\emph{Suzaku}$ data with the EPIC-pn data provides a better estimate of the physics occurring within the source. The EPIC-pn spectrum has a limited bandwidth which may not render the proper spectral index. In order to obtain a better estimate of the photon index, $\Gamma$, we fit a grid of {\sc relxill} models (for $a_{*}=0,0.5,0.99$ and $i<45$) with a blackbody disk component to the $\emph{RXTE}$ spectra for the simultaneous observation taken in 2002. A constant was allowed to float between the PCA and HEXTE. The photon index, $\Gamma$, was determined to be $2.56_{-0.03}^{+0.04}$, $2.61\pm0.03$, and $2.62_{-0.05}^{+0.06}$ for the different values of fixed spin. To account for calibration uncertainties between the two missions, we restrict the photon index of the EPIC-pn spectrum to be within $\Delta\Gamma$=0.1 of the average value of $\Gamma=2.60$ when applying models. Again, the $\emph{XMM-Newton}$ data were reduced two different ways with and without the task {\sc epfast}. We initially used the data that were reduced with {\sc epfast} when modeling the data jointly with $\emph{Suzaku}$. We then overlaid the spectrum that was generated without {\sc epfast} onto the best fit to see if there was a significant difference between the two $\emph{XMM-Newton}$ spectra. The differences were very small so we chose to use the data generated with the task {\sc epfast}. We ignore the 1.75-2.35 keV range in the $\emph{XMM-Newton}$ data rather than fitting multiple Gaussians to account for features of instrumental origin as per \citet{Plant15}. Additionally, we ignore below 1.3 keV due to residuals that were also identified by \citet{Reis11}, \citet{hiemstra}, and \citet{Plant15}. For the $\emph{Suzaku}$ data, we ignore the 1.5-1.9 keV and 2.1-2.4 keV energy ranges due to the poorly calibrated known Si-K and Au-M instrumental residuals.

As stated before, the model that we use to fit the spectra is {\sc tbabs*(diskbb+relxill)}. 
In Table 1 are the values obtained when performing a joint fit between the $\emph{XMM-Newton}$ and $\emph{Suzaku}$ observation in the 1.0-10.0 keV energy range and extended out into the higher energy range from 15.0-60.0 keV with the spectra obtained from the HXD/PIN. The spin and inclination have been tied between the two observations because those should remain consistent between observations and state changes.  The angleon parameter can be set to either 0 or 1. Setting the angleon parameter to 0 causes the model to angle-average the X-ray reflection. Using an angle-averaged X-ray reflection model can bias the inclination to lower values \citep{garcia14}. We make sure to set the angleon=1 so that the inclination is properly taken into account and not angle averaged.  Figure 1 shows the simultaneous modeling of the spectra from $\emph{XMM-Newton}$ and $\emph{Suzaku}$. Figure 2 shows the unfolded model spectrum for the  $\emph{Suzaku}$ XIS and HXD/PIN as reference for what the models are accounting for the data.

We also assume in this model that the inner accretion disk is fully extended to the ISCO. Disk truncation is still a topic of debate, though there has been some recent evidence presented for truncation in the intermediate state (\citealt{tam12}, \citealt{Plant15}), it seems to be unlikely even in the low/hard states (see \citealt{miller06}; \citealt{rykoff}; \citealt{reis09}, \citeyear{reis10}). A strong indicator that the inner disk is extended down to the ISCO is the presence of high frequency quasi-periodic oscillations (\citealt{nowak00}, \citealt{Nespoli03}) in the intermediate state and the VHS. The frequencies are proportional with the orbits close to the ISCO (\citealt{Stroh01}, \citealt{miller01}, \citealt{homan03}, \citeyear{homan05b}). We allow the emissivity, $q$, to be broken since constant emissivity throughout the disk is often thought to not be physically reasonable due to the difference in conditions in the accreting material closer to the BH \citep{Wilkins12}.  For this reason we then allowed the inner emissivity ($q_{in}$) to vary between 3-10 $R_{g}$ and the outer emissivity index ($q_{out}$) to vary between 0-3 $R_{g}$. The break radius was restricted to lie between 3-6 $R_{g}$. 

It can be seen that the model returns an acceptable fit to the data ($\chi^{2}/dof = 1816.36/1472$). The high ionization parameter is consistent with those obtained in \citet{reis08} when looking at observations taken in the VHS and LHS. It was found that ionization parameter for the VHS was $log(\xi) > 4$ and in the LHS to be $log(\xi) \approx 3$. It makes sense that the ionization parameter in the intermediate state would fall between the two yet be closer in value to the VHS since it was taken after the outburst had peaked. The ionization parameter value for the $\emph{XMM-Newton}$ spectrum is consistent with the hard limit of 6 within the 90\% confidence level. The iron abundance was allowed to be free and returns a value that is $\approx 4$ times the solar abundance. This may be an overestimate that can be attributed to the {\sc{relxill}} model.  See \citealt{garcia14} and \citealt{kara15} for further explanation. The inner emissivity for the  $\emph{Suzaku}$ spectra is not well constrained but it is clearly not driving the spin result. The temperature of the disk is consistent with those found in \citet{miller08}. The high reflection fraction suggest that the spectra is reflection dominated, but this is indicative of a BH with a high spin value (\citeauthor{dauser14} \citeyear{dauser14}, \citeauthor{Parker} \citeyear{Parker}).  The constant between the HXD/PIN and XIS is low with respect to the standard $\sim1.16$\footnote{JX-ISAS-SUZAKU-MEMO-2008-06 by Maeda et al.}, but this could be from extracting an annular region and having used an uncommon XIS mode. 

We find the  inclination of GX 339-4 to  be $36 \pm 4$ degrees. Using the mass function of $5.8\pm0.5$ M$_\odot$ \citep{hynes03} and a mass fraction of 0.125 \citep{Munoz08} for this binary system with the inclination above suggests a large BH  mass of $\sim 37.5$ M$_{\odot}$. This applies if and only if the inclination of the binary system and the BH are aligned, but this need not be the case. A lower mass estimate can be achieved if the the inclination of the BH and the binary are misaligned. The true inclination is hard to determine as discussed earlier in this section. A higher inclination would also return a lower BH mass. 

We obtain a spin parameter of $a_{*}$ $>0.97$. This confirms the high spin estimates that have been found for GX 339-4 in the previous years. The maximum value for spin allowed for a BH is 0.998 \citep{Thorne}. When constraining the error on spin we find that the maximum allowed value is within the 90\% confidence level. We therefore quote the difference between the maximum allowed value and value found for spin as the upper bound. Figure 3 shows the change in $\chi^2$ when stepping the spin parameter from 0 to 0.998 using the \lq \lq steppar" command in XSPEC. 

\section{Discussion}
We have found the spin of GX 339-4 to be a near maximally spinning BH with a spin parameter of  $a_{*}$ $>0.97$. The high reflection fraction also indicates a rapidly spinning BH (\citeauthor{dauser14} \citeyear{dauser14}, \citeauthor{Parker} \citeyear{Parker}). This is consistent previous high spin estimates made by \citet{miller08} and \citet{reis08}. 
 The analysis made in previous works and our analysis performed with {\sc relxill} are different in several respects:
\begin{itemize}
\item  {\sc refhiden} allows for a different atmosphere, heated by a strong blackbody in the mid plane.
\item {\sc relxill} has a higher spectral resolution, improved atomic data, and includes more lines than prior models. 
\item {\sc relxill} can be set to take the inclination angle into account when modeling X-ray reflection, whereas {\sc reflionx} and {\sc refhiden} are angle-averaged.
\end{itemize}
It is important to account for systematic errors. Systematics owing to any violation of the test particle ISCO by an actual fluid disk are likely important, but appear to be small in recent numerical simulations \citep{Reynolds08}.  We have found that the combination of atmospheric structure, model resolution, and inclination angle - each potentially an important source of systematic error - only produce $\sim$5\% offsets in combination. In other words, they are not large sources of systematic error.   It should be noted that inclination values that would keep the mass of the black hole in GX 339$-$4 below 10~$M_{\odot}$ would necessitate eclipses, which are clearly not seen.

\begin{figure}
\centering
\includegraphics[width=8.4cm]{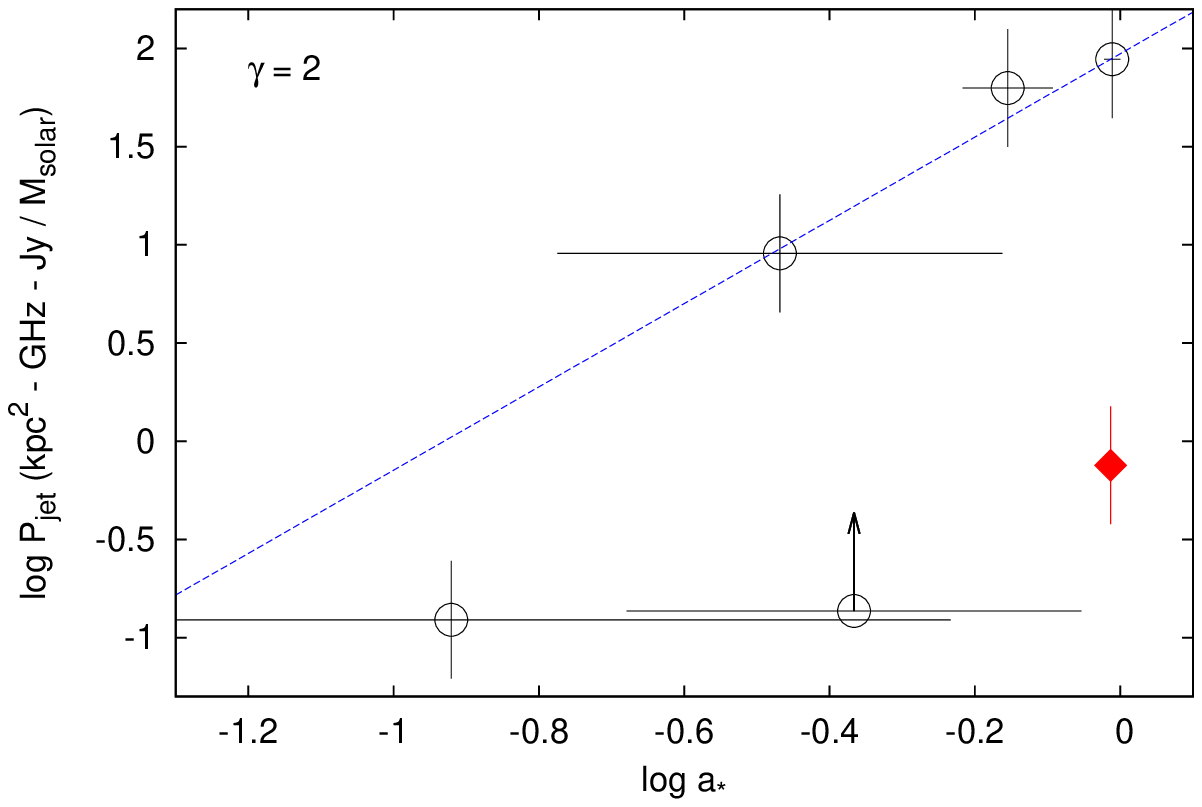}
\includegraphics[width=8.4cm]{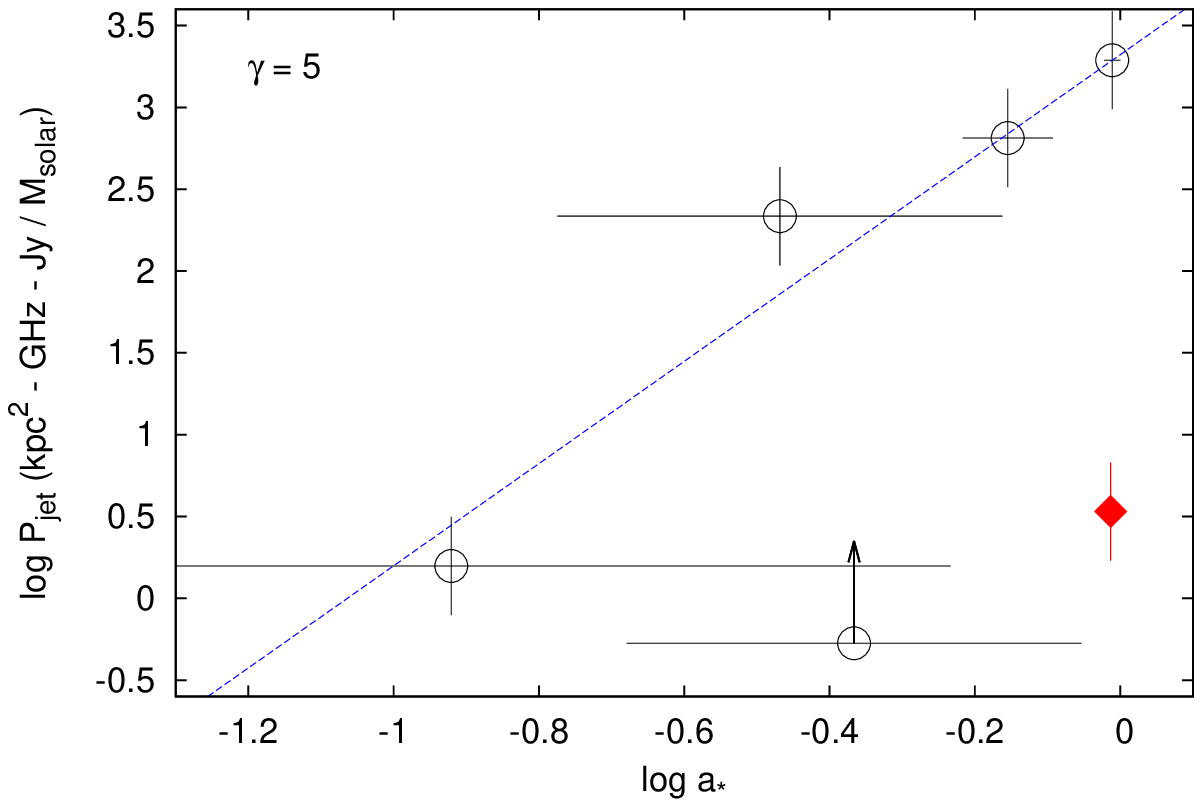}
\caption{Black hole spin versus jet power. See Table 2 for values. GX 339-4 is given in red. Errors are plotted for the spin of GX 339-4 though they are to small to see. The top figure is for $\gamma = 2$. The slope of the line is $2.12 \pm 0.75$ which is close to that predicted for the BZ effect. The lower figure is for $\gamma = 5$ and has a slope of $3.12 \pm 0.39$.}
\label{fig:spinpower}
\end{figure} 

Given that GX339-4 is an important source, and that our results again point to a very high BH spin value, it is worth considering the role of spin on GX 339-4 and other stellar-mass BHs. In particular, we look into where GX 339-4 falls according to existing relations between BH spin and jet power pertaining to BH binary (BHB) systems. One such relation in \citet{Narayan12} has found a near linear correlation for five BHBs with low mass companions and that produce ballistic jets during outburst. This appears to confirm the predictions of \citet{BZ} (BZ) where the relativistic jets are being powered by extracting energy from the angular momentum of the BH, though it was \citet{penrose69} who first theorized this concept.  \citeauthor{Narayan12} do not include GX 339-4 in their correlation, but provide a rough estimate of the spin using their correlation. They estimate GX 339-4 to have a spin between 0.2-0.5 when assuming that the maximum observed radio flux is 55 mJy \citep{gallo04}, which differs from our findings significantly. We take the observed flux density (at 5 GHz) and spin values that are presented in \citet{Narayan12} for four of the five sources: A0620-00, XTE J1550-564, GRO J1655-40, GRS 1915+105. These spins were determined via the continuum fitting method. For the maximum observed radio flux density of GX 339-4 we use 55 mJy \citep{gallo04}. Recently, \citet{morningstar14} conducted a spin estimate for 4U 1543-47 using the iron line emission similar to what we have done here for GX 339-4. We choose to use this recent estimate when looking into the relation between spin and jet power. The radio flux of 4U 1543-47 is seen as a lower limit for the jet power and is indicated as such in Figure 4 since it lacks sufficient radio data throughout the entirety of its outbursts \citep{park}. 

To change the observed fluxes into jet power we have to correct the fluxes for Doppler boosting. We use an equation from \citet{Mirabel} to correct for this effect.
\small
\begin{equation}
S_{cor} = \frac{S_{obs}}{ \delta^{3-\alpha}}
\end{equation}
\normalsize

where the Doppler factor $\delta = (\gamma(1-\beta \cos \theta))^{-1}$,  $\gamma = (\sqrt{1-\beta^{2}})^{-1}$ is the Lorentz factor, $\beta = v/c$, and $\alpha$ is the radio spectral index. We use updated values for $\alpha$ from \citet{king13a}. These calculations were performed for two different values of Lorentz factors ($\gamma = 2$ \& $\gamma = 5$). $\gamma = 2$ is chosen analogous with the work done in \citet{Narayan12}, but since the Lorentz factor is difficult to determine we choose a higher value of $\gamma = 5$ for comparison. We can then use the corrected flux for each object to estimate to the power of the jet being emitted from the BH by using the relation given \citet{Narayan12}.

\small
\begin{equation}
P_{jet} = D^{2} \nu S_{\nu} / M_{BH}
\end{equation}
\normalsize

\begin{table}
\caption{Parameters for BH spin and Jet Power}
\label{tab:comp}
\begin{center}
\begin{tabular}{lcccc}
\hline
\multicolumn{1}{c}{BH Binary}
&\multicolumn{1}{c}{$a_{*}$}
&\multicolumn{1}{c}{$P_{jet} (\gamma = 2)$}
&\multicolumn{1}{c}{$P_{jet} (\gamma = 5)$}
&\multicolumn{1}{c}{ref}
\\
\hline
A0620-00
&$0.12 \pm 0.19$
&$0.12$
&$1.6$
&1 \\
XTE J1550-564
&$0.34 \pm 0.24$
&$9.1$
&$220$ 
&1\\
GRO J1655-40
&$0.7 \pm 0.1$
&$63$
&$650$
&1\\
GRS 1915+105
&$0.975 \pm 0.025$
&$88$
&$1900$
&1\\
4U 1543-47
&$0.43 \pm 0.31$
&$0.14$
&$0.53$
&2\\
GX 339-4
& $>0.97$
&0.75
&3.4
&3\\
\hline
\end{tabular}

\medskip
Note.--- The units of $P_{jet}$ are given in kpc$^2$ GHz Jy M$_{\odot}^{-1}$. The reference refers to where the spin and/or observed maximum radio flux was taken. (1) \citeauthor{Narayan12} \citeyear{Narayan12} (2) \citeauthor{morningstar14} \citeyear{morningstar14}  (3) \citeauthor{gallo04} \citeyear{gallo04}.
\end{center}
\end{table}

Where $D$ is the distance to the object, $\nu$ is the observation frequency (taken to be 5 GHz), $S_{\nu}$ is the corrected radio flux density, and $M_{BH}$ is the mass of the BH. For GX 339-4 we use $D=8$ kpc \citep{zd04} with $M_{BH}=5.8\pm0.5 M_\odot$ \citep{hynes03}. The values for jet power can be seen in Table 2. Figure 4 shows the jet power in relation to the spin of the BHB on a logarithmic scale. The slope of weighted linear regression line for $\gamma = 2$ is  $2.12 \pm 0.75$, which is roughly consistent with that predicted of the BZ effect. However, the Lorentz factor is very difficult to constrain in many systems and is therefore largely uncertain \citep{fender03}. When we allow the Lorentz factor to increase to 5, the slope of the best fit line becomes $3.12 \pm 0.39$. Yet, this steeper slope is still within $3\sigma$ from the value at $\gamma = 2$. This is not highly significant and can still be considered consistent with the BZ effect.

It can be seen in either case that GX 339-4 lies significantly below the predicted jet power via spin energy extraction for a near maximally spinning BH. If we use the higher mass estimate of $\sim 37.5$ M$_{\odot}$, when assuming the the inclination of the binary and BH are aligned, then the source is positioned even further away the from the correlation between $a_{*}$ and $P_{jet}$. One explanation for the poor fit could be that the jet power is not correlated with the spin alone. \citet{Fender10} and \citet{Russell13} have found that spin and jet power are not correlated when a larger sample is used. \citeauthor{king13b} (2013a,b) have investigated the role of spin in driving jets across the BH mass scale. They find that spin may set the maximum power in the jet, but that the mass accretion rate may act as a \lq throttle' in setting the instantaneous jet power, as measured via radio flux. Both \citet{Russell13} and \citet{king13a} have found that a positive correlation between $P_{jet}$ and $a_{*}$ is, at best, marginally significant for BHXBs ($\leq 2.3 \sigma$ confidence level). Our results are consistent with these findings. It is clear that the instantaneous jet power in GX 339-4 is not set by spin. 

We would like to thank the anonymous referee for the helpful and constructive report. This research has made use of data and/or software provided by the High Energy Astrophysics Science Archive Research Center (HEASARC), which is a service of the Astrophysics Science Division at NASA/GSFC and the High Energy Astrophysics Division of the Smithsonian Astrophysical Observatory.

\bibliographystyle{apj}
\bibliography{apj-jour,references}

\end{document}